\documentclass[iop,apj]{emulateapj}
\usepackage{times}
\usepackage{epsfig}
\usepackage{amsmath, amsthm, amssymb}
\usepackage[plainpages=false, colorlinks=true, anchorcolor=blue, linkcolor=blue, citecolor=blue, bookmarks=false]{hyperref}
\usepackage{color}
\usepackage{microtype}
\usepackage{float}
\usepackage{verbatim}

\newcommand{\beq}{\begin{equation}}
\newcommand{\eeq}{\end{equation}}
\newcommand{\bea}{\begin{eqnarray}}
\newcommand{\eea}{\end{eqnarray}}
\newcommand{\Msun}{$M_\sun$}

\newcommand{\lmax}{$l_{\rm max}$}

\slugcomment{Submitted to the Astrophysical Journal}

\begin{document}

\title{An Improved Multipole Approximation for Self-Gravity and Its
  Importance for Core-Collapse Supernova Simulations}
\author{Sean M. Couch\altaffilmark{1}, Carlo Graziani, \& Norbert Flocke}

\affil{Flash Center for Computational Science, Department of Astronomy
  \& Astrophysics, University of Chicago, Chicago,
IL, 
60637; smc@flash.uchicago.edu}
\altaffiltext{1}{Hubble Fellow}

\shorttitle{MULTIPOLE GRAVITY}
\shortauthors{COUCH, GRAZIANI, \& FLOCKE}

\begin{abstract}
  Self-gravity computation by multipole expansion is a common approach
  in problems such as core-collapse and Type Ia supernovae, where
  single large condensations of mass must be treated.  The standard
  formulation of multipole self-gravity in arbitrary coordinate systems suffers from two significant
  sources of error, which we correct in the formulation presented in
  this article.  The first source of error is due to the numerical
  approximation that effectively places grid cell mass at the central
  point of the cell, then computes the gravitational potential at that
  point, resulting in a convergence failure of the multipole
  expansion. We describe a new scheme that avoids this problem by
  computing gravitational potential at cell faces. The second source
  of error is due to sub-optimal choice of location for the expansion
  center, which results in angular power at high multipole $l$ values
  in the gravitational field, requiring a high --- and expensive ---
  value of multipole cutoff \lmax.  By introducing a global measure of
  angular power in the gravitational field, we show that the optimal
  coordinate for the expansion is the square-density-weighted mean
  location.  We subject our new multipole self-gravity algorithm, implemented in the FLASH simulation framework, to
  two rigorous test problems: MacLaurin spheroids for which exact
  analytic solutions are known, and core-collapse supernovae.  We show
  that key observables of the core-collapse simulations, particularly
  shock expansion, proto-neutron star motion, and momentum
  conservation, are extremely sensitive to the accuracy of the
  multipole gravity, and the accuracy of their computation is greatly
  improved by our reformulated solver.


\end{abstract}

\section{Introduction}
\label{sec:Intro}

Gravity is a key phenomenon in many astrophysical contexts, and, in
particular, plays an essential role in the explosions of core-collapse
supernovae (CCSNe). Accurate computation of self-gravity is therefore
an important objective for astrophysical simulation codes.  For
self-gravitating Newtonian systems this requires solving Poisson's
equation.  Poisson's equation is an elliptic partial differential equation, which couples every
part of the domain at each time step.  The optimal solver strategy for
an astrophysical Poisson problem in which gravity is coupled to a
hydrodynamic flow depends on the typical mass configuration in the
domain.  Multigrid algorithms \citep[e.g.,][]{{Huang:2000vd},
  trottenberg2001, Ricker:2008eq} are popular for cosmological
structure-formation and star formation simulations with Newtonian
gravity \citep[e.g.][]{Yang:2009iy, ZuHone:2010ev, Latif:2011ir,
  Federrath:2012em}, since these algorithms work well with mass
configurations spread out over a computational domain.  In problems
where a single, large condensation of mass arises, however, a
multipole expansion using spherical harmonics is more appropriate.
Solving Poisson's equation using spherical harmonic expansions is a
common approach for computing the self-gravity of nearly-spherical
mass distributions. Multipole approximations have been used in a
number of astrophysical applications including N-body calculations
\citep[see][]{Sellwood:1987ih} and grid-based hydrodynamics
\citep[][]{Muller:1995jr}.  The unstable collapse of the core of
a massive star that preceeds a core-collapse supernova (CCSN) is
particularly sensitive to a highly dynamic gravitational potential.
Many approaches have been adopted for treating self-gravity in CCSN
simulations ranging from full general relativity
\citep[e.g.,][]{OConnor:2010bi, Ott:2007cc, 2010ApJS..189..104M,
  {Ott:2013gz}} to simplified 1D ``monopole'' approximations
\citep[e.g.,][]{Hanke:2012dx,Dolence:2013iw,Couch:2012un}.

The Newtonian potential of a spherically-symmetric self-gravitating
mass is trivial, of course, and is represented by the monopole term of
the expansion.  However, as departures from spherical symmetry
accumulate, the mass distribution must be represented by an expansion
of spherical harmonics beyond $l=0$, the accuracy of which depends on
the degree of non-sphericity of the mass distribution and the number
of terms used in the expansion \citep{Muller:1995jr}.  Such multipole
approaches for self-gravity have been used in a number of
multidimensional CCSN simulations \citep[e.g.,][]{2004ApJ...609..277L,
 {Buras:2006dl},{Bruenn:2013es}}.  Multipole approaches are suited for CCSNe because
the gravitational potential is dominated by the monopole contribution,
but the higher-order contributions due to significant non-spherical
motions in the post-shock region can be important.  Additionally, in
non-spherical geometries wherein the central proto-neutron star is
allowed to move, the physical kick imparted on the star by the
requirement of momentum conservation --- a model-constraining
observable --- is critically dependend on an accurate,
momentum-conserving self-gravity computation
\citep{Wongwathanarat:2010ch, {Wongwathanarat:2012tf}}.

In this article, we investigate the multipole expansion approach to solving 
Poisson's equation numerically for the self-gravity of an approximately 
spherical mass distribution.  We identify, and correct, two heretofore 
neglected sources of significant errors that arise in 
implementations of multipole self-gravity for non-spherical coordinates: 

\begin{enumerate} 

\item\label{item:cellface} The numerical approximation that effectively 
places grid cell mass at the central point of a computational cell, then 
computes the gravitational potential at that point, resulting in a 
convergence failure of the multipole expansion, so that larger choices of 
multipole cutoff value \lmax\ actually make the potential computation 
less accurate;

\item\label{item:center} Sub-optimal choice of location for the expansion 
center, which results in angular power at high multipole $l$ values in the 
gravitational field, requiring a high --- and expensive --- value of 
\lmax;
  

\end{enumerate}

We show here that source \ref{item:cellface} of error can be
eliminated by a collocation scheme that effectvely staggers point mass
placement and potential computation; and source \ref{item:center} of error
can be minimized by a careful, unique choice of expansion center,
which we derive.  We demonstrate that CCSN simulations are
particularly sensitive to these details of the multipole approximation
for gravity and show that our improvements result in dramatic
improvements in important metrics such as momentum conservation and
convergence with number of terms in the multipole expansion.

This paper is organized as follows.  In \S\ref{subsec:SelfPotError} we
briefly present the discretized multipole equations and exhibit the
intrinsic error that can afflict solutions to these equations due to
the singularity in the Green's function of Poisson's equation.  We
show that this error is eliminated by computing gravitational
potentials on cell faces rather than at cell centers.  In
\S\ref{subsec:Centering} we derive the optimal location for centering
the expansion for general mass distributions, based on the
minimization of an angular ``spectral compactness'' measure that
characterizes the extent in $l$-space of the spherical-harmonic
spectrum.  In Section \ref{sec:implement} we describe our
implementation of fast, efficient multipole gravity solver in the
FLASH simulation framework.  We test our new solver, which includes
the improvements we discuss, on static potentials in Section
\ref{sec:SimulationMacLaurin} and exhibit the effects of the errors
described above, as well as the result of their correction.  In
Section \ref{sec:ccsn} we test our new implementation with highly
dynamical CCSN simulations in two dimensions and show that the results
are highly sensitive to the centering of the multipole expansion and
to the relative collocation of the mass and potential evaluation
points.  We discuss our conclusions in Section \ref{sec:conclusions}.


\section{Discretized Multipole Expansions}
\label{sec:mpoleExp}
\subsection{The Self-Potential Error\label{subsec:SelfPotError}}

The gravitational potential of an isolated distribution of mass with
density $\rho(\mathbf{x})$ is given by the well-known Green's function
of the Poisson equation 
\begin{equation}
  \Phi(\mathbf{x})=-G\int d^{3}\mathbf{x}^{\prime}\frac{\rho(\mathbf{x}^{\prime})}{\left|\mathbf{x}-\mathbf{x}^{\prime}
  \right|}.\label{eq:Potential}
\end{equation}
Direct numerical implementation of this formula in a simulation is
inefficient, often necessitating approximate approaches.  For mass
distributions that can be described as spherical to lowest-order,
multipole expansions of Equation (\ref{eq:Potential}) can be used to
efficiently compute solutions.  The multipole expansion version of the
potential is given by the equally well-known formula
\begin{eqnarray}
  \Phi(\mathbf{x})&=&-G\sum_{l=0}^{\infty}\sum_{m=-l}^{l}\frac{4\pi}{2l+1} \nonumber
  \\ &\times& \int d^{3}\mathbf{x}^{\prime}\,\rho(\mathbf{x}
  ^{\prime})\, Y_{lm}(\mathbf{n})Y_{lm}(\mathbf{n}^{\prime})^{*}g_{l}(r,r^{\prime}),\label{eq:Potential_Expansion}
\end{eqnarray}
where $r\equiv|\mathbf{x}|$, $\mathbf{n}\equiv\mathbf{x}/|\mathbf{x}|$,
and\textbf{ }
\begin{equation}
  g_{l}(r,r^{\prime})\equiv\Theta(r-r^{\prime})\frac{r^{\prime l}}{r^{l+1}}+\Theta(r^{\prime}-r)\frac{r^{l}}{r^{\prime l
      +1}}.\label{eq:g_factor}
\end{equation}
Here, $\Theta(x)$ is the usual Heaviside function.

In Eulerian hydrodynamic codes, a standard discretization strategy
for this expansion \citep{Muller:1995jr} begins with a
subdivision of the domain into $N_{R}$ spherical shells bounded by
radii $R_{t},$ $t=1,\ldots,N_{R}$, chosen to suit the problem (and
not necessarily uniformly spaced). A cell centered at the position
$\mathbf{x}_{q}$ is ascribed a radius $r{}_{q}$ that is the mean
radius of the spherical shell containing $\mathbf{x}_{q}$, where
$q$ is an index running over mesh cells. The discretized potential
is then computed as
\begin{eqnarray}
  \Phi(\mathbf{x}_{q})&=&-G\sum_{l=0}^{l_{\rm
      max}}\sum_{m=-l}^{l}\frac{4\pi}{2l+1}Y_{lm}(\mathbf{n}_{q})\nonumber 
  \\ &\times& \sum_{q^{\prime}}
  \Delta_{q^{\prime}}^{3}\rho(\mathbf{x}_{q^{\prime}})Y_{lm}(\mathbf{n}_{q^{\prime}}^{\prime})^{*}
  \nonumber
  \\ &\times& \left\{ \tilde{\Theta}
    _{qq^{\prime}}\frac{r_{q^{\prime}}^{l}}{r_{q}^{l+1}}+
    \tilde{\Theta}_{q^{\prime}q}\frac{r_{q}^{l}}{r_{q^{\prime}}^{l+1}}+
    \tilde{\delta}_{qq^{\prime}}\frac{1}{r_{q}}\right\} ,\label{eq:Multipole_Discretized}
\end{eqnarray}
where $l_{\rm max}$ is some chosen cutoff value for the expansion,
$\Delta_{q^{\prime}}^{3}$
is the volume of the cell centered at $\mathbf{x}_{q^{\prime}}$,
and where
\[
\tilde{\Theta}_{qq^{\prime}}\equiv\begin{cases}
  1 & r_{q}>r_{q^{\prime}}\\
  0 & r_{q}\le r_{q^{\prime}}
\end{cases}\quad;\quad\tilde{\delta}_{qq^{\prime}}\equiv\begin{cases}
  1 & r_{q}=r_{q^{\prime}}\\
  0 & r_{q}\ne r_{q^{\prime}}
\end{cases}.
\]

If one were directly implementing the potential using the expression
of Eq.~(\ref{eq:Potential}), discretization in the presence of the
singular Green's function
$\left|\mathbf{x}-\mathbf{x}^{\prime}\right|^{-1}$ might give rise to
misgivings having to do with the delicate handling of gravitational
self-interaction within a mesh cell. This issue of self-gravity
appears superficially to magically cure itself in the passage to the
discrete multipole expansion of Equation (\ref{eq:Multipole_Discretized}),
wherein no short-distance singularities are explicitly visible. This
miracle cure is illusory, unfortunately: the singularity still lurks
in the expression, and manifests itself in the failure of the
self-interaction terms in the expression to converge as $l_{\rm
  max}\rightarrow\infty$.

To see this, consider the self-interacting term $q^{\prime}=q$ in
Equation (\ref{eq:Multipole_Discretized}):
\begin{eqnarray}
  \Phi_{\rm Self}(\mathbf{x}_{q}) &\equiv& 
  \frac{-G\Delta_{q}^{3}\rho(\mathbf{x}_{q})}{r_{q}}\sum_{l=0}^{l_{\rm max}}\frac{4\pi}
  {2l+1} \nonumber \\ 
  &\times& \sum_{m=-l}^{l}Y_{lm}(\mathbf{n}_{q})Y_{lm}(\mathbf{n}_{q})^{*}.\label{eq:Phi_Self_1}
\end{eqnarray}
The addition theorem of spherical harmonics states that
\begin{equation}
  \sum_{m=-l}^{l}Y_{lm}(\mathbf{n}_{1})Y_{lm}(\mathbf{n}_{2})^{*}=
  \frac{2l+1}{4\pi}P_{l}(\mathbf{n}_{1}\cdot\mathbf{n}_{2}),
  \label{eq:Addition_Theorem}
\end{equation}
where $P_{l}(x)$ is a Legendre polynomial. We therefore have that
\begin{eqnarray}
  \Phi_{\rm Self}(\mathbf{x}_{q}) & \equiv & 
  \frac{-G\Delta_{q}^{3}\rho(\mathbf{x}_{q})}{r_{q}}\sum_{l=0}^{l_{\rm max}}P_{l}
  (1)\nonumber \\
  & = & \frac{-G\Delta_{q}^{3}\rho(\mathbf{x}_{q})}{r_{q}}\sum_{l=0}^{l_{\rm max}}1\nonumber \\
  & = &
  \frac{-G\Delta_{q}^{3}\rho(\mathbf{x}_{q})}{r_{q}}\times(l_{\rm
    max}+1).
  \label{eq:Self_Energy_Divergence}
\end{eqnarray}
It follows that the discrete expression for $\Phi_{\rm
  Self}(\mathbf{x}_{q})$ is not convergent with multipole order, and
that the accuracy of the discrete scheme described above cannot be
improved by increasing $l_{\rm max}$.  We note that
\citet{Sellwood:1987ih} remarked upon related difficulties in the
context of $N$-body simulations, but did not give the explicit form of
this self-potential error nor expound on its origins in the discrete
multipole expansion.  In numerical simulations, this pathology
manifests itself as a dramatic failure in accuracy of the potential
calculation, which gets worse with increasing $l_{\rm max}$.  We also
note that due to the factor $r_q$ in the denominator of
Equation (\ref{eq:Self_Energy_Divergence}) this error is worst near the
origin of the multipole expansion.  This error is also larger for
computational zones containing large masses,
$\Delta_{q}^{3}\rho(\mathbf{x}_{q})$.  Both of this conditions are met
in the extreme for CCSN simulations containing a proto-neutron star
near the center of the domain.

A more deft handling of self-interaction is required if the scheme is
to be rescued. We may begin by observing that the physical origin of
the difficulty is that the scheme in effect treats all masses as
points at the cell centers, then computes potentials at those same
cell centers. If the points of potential computation were offset from
the cell centers, the problem would go away. This is akin to the idea
of a gravitational softening length.  Mathematically, in the
limit $l_{\rm max}\rightarrow\infty$, the self-gravity expression
calculated at an offset point $\mathbf{x}_{\rm off}$ near
$\mathbf{x}_{q}$ is
\begin{eqnarray}
\Phi_{\rm Self}(\mathbf{x}_{\rm off}) & = & \frac{-G\Delta_{q}^{3}\rho(\mathbf{x}_{q})}{r_{q}}\sum_{l=0}^{\infty}P_{l}
(\mathbf{n}_{\rm off}\cdot\mathbf{n}_{q})\nonumber \\
 & = & \frac{-G\Delta_{q}^{3}\rho(\mathbf{x}_{q})}{r_{q}}\left[2\left(1-\mathbf{n}_{\rm off}\cdot\mathbf{n}_{q}\right)
\right]^{-1/2},\label{eq:Self_Energy_Finite}
\end{eqnarray}
where we have used the generating function of the Legendre polynomials,
$(1-2xt+t^{2})^{-1/2}=\sum_{l=0}^{\infty}t^{l}P_{l}(x)$
\citep{Arfken:2005uf} with $t=1$. This expression is obviously finite, so the
expansion converges. Of course, we need the potential at cell centers
to compute gravitational forces --- momentum and energy fluxes --- at cell faces.
So we modify the basic scheme above by computing potentials at all
cell faces, and ascribing to each cell center the average of the potentials
on the faces bounding the cell. This should be a very accurate
operation as the gravitational potential is generally a smooth
function in space.  As shown below, this scheme works
well: it converges with multipole order, and provides excellent momentum
conservation. 

It is important to note that the self-potential error described above is a product of the discrete evaluation of Equation (\ref{eq:Potential_Expansion}).
In spherical coordinates, it is possible to compute Equation (\ref{eq:Potential_Expansion}) {\it analytically}, assuming constant density within the zone \citep{Muller:1995jr}.
Such an approach is not subject to the self-potential error (A. Wongwathanarat 2013, private communication).
Analytic evaluation of Equation (\ref{eq:Potential_Expansion}) in general coordinate systems is more difficult, particularly in non-spherical curvilinear systems.
Thus, in order to retain uniformity amongst different coordinate systems while avoiding the self-potential error, we choose to evaluate the potentials discretely at cell faces, as discussed above. 

\subsection{Optimal Centering of a Multipole Expansion\label{subsec:Centering}}

The issue of where a multipole expansion should be centered has received
surprisingly little analytic attention, given its importance to accurate
computation of the gravitational potential. A possible reason for
this is that in many cases, a spherical coordinate system is adopted,
obviating the ambiguity in the choice of expansion center.  For
other coordinate geometries, such as cylindrical and Cartesian, the
optimal location of the expansion origin is not so obvious, and a
careless choice can be costly to the accuracy of the gravity solve.

There exist intuitive arguments for different choices of expansion
center. The center of the grid is the obvious choice in spherical
coordinate meshes. The center-of-mass (CoM) is indicated, perhaps a
little indirectly, on the basis of the importance that it plays as a
diagnostic of linear momentum conservation, since motion of the CoM
directly indicates a failure of momentum conservation.  The CoM is
also a good choice as centering the expansion there eliminates the
$l=1$ dipole term
\citep[e.g.,][]{Muller:1995jr}. \citet{McGlynn:1984cp} working in an
$N$-body context, advocates an expansion center location $\mathbf{a}$
minimizing the sum $\sum_{n}|\mathbf{x}_{n}-\mathbf{a}|^{2k}$, with
the parameter $k$ chosen empirically to balance the relative weighting
of inner and outer particles. 
\citet{McGlynn:1984cp} also points out
that the truncated multipole expansion is not translationally invariant, a point
that has significant consequence for the conservation of linear
momentum in calculations relying on multipole gravity solvers.  This
feature of multipole expansions underscores the criticality of
optimally centering the expansion so as to best maintain momentum
conservation. 

\citet{Sellwood:1987ih} stresses that the origin of the multipole
expansion should be placed at the location of peak density, because failure
to do so can result in errors in the gravitational force,
and in energy non-conservation.  The intuitive reason that the
peak density makes sense as the expansion origin is that condensations
at large radii subtend small angles at the origin, and, if massive, can show up
as power in higher-$l$ regions of the angular momentum spectrum than
would be the case were they placed near the center. It is important
that the angular power spectrum of the potential be concentrated to as
low values of $l$ as is practicable, because discrete multipole
Poisson solvers truncate the expansion in spherical harmonics at some
$l_{\rm max}$. This cutoff should be as low as possible, for the sake
of computational efficiency [the computational cost of the Poisson
solve grows as $\mathcal{O}(l_{\rm max}{}^{2})$ in three dimensions],
but higher than any substantial power in the spectrum.

In this section we give more rigorous arguments than have been offered
to date for the choice of expansion center. We use angular spectral
``compactness'', as described informally above, as the criterion for
making the choice. We show that the choice advocated by
\citet{Sellwood:1987ih} is, for all intents and purposes, very close
to optimal when there is a significant fraction of total mass in a
condensed object.

\subsubsection{Spectral Compactness Minimization\label{subsection:Compactness}}

As adumbrated above, we need a way to characterize the global angular spectral
distribution in the gravitational field, so as to have some way to discuss how
well the spectrum is concentrated to low values of $l$. 

The multipole expansion of the potential $\Phi(\mathbf{x})$, given in
Eqs.~(\ref{eq:Potential_Expansion}) and (\ref{eq:g_factor}), is not ideal for this 
purpose, since its spectral content varies in space. We may, however, 
average $\Phi(\mathbf{x})$ spatially, weighted by the density 
$\rho(\mathbf{x})$, to obtain the binding energy,
\begin{eqnarray}
  \mathcal{E} & = & -\frac{1}{2}\int d^{3}\mathbf{x}\,\Phi(\mathbf{x})\rho(\mathbf{x})\nonumber \\
  & = & \sum_{l=0}^{\infty}\mathcal{E}_{l},\label{eq:Binding_Energy}
\end{eqnarray}
where
\begin{eqnarray}
  \mathcal{E}_{l} & \equiv & \frac{G}{2}\frac{4\pi}{2l+1}\times\nonumber\\
  &&\sum_{m=-l}^{l}\int d^{3}\mathbf{x}\, d^{3}\mathbf{x}^{\prime}\,
  \rho(\mathbf{x})\rho(\mathbf{x}^{\prime})\, Y_{lm}(\mathbf{n})Y_{lm}(\mathbf{n}^{\prime})^{*}g_{l}(r,r^{\prime}),\nonumber\\
\label{eq:E_l_1}
\end{eqnarray}
and $g_{l}(r,r^{\prime})$ is the function given in Eq.~(\ref{eq:g_factor}).

We propose to use $f_l\equiv\mathcal{E}_{l}/\mathcal{E}$ as a global angular
spectral density in what follows. In order for this to make sense,
it is of course necessary to establish that $\mathcal{E}_{l}\ge0$ for all $l$.
We demonstrate that this is the case in the Appendix.

How can we measure the concentration to low $l$ of the
distribution $f_{l}$?  A reasonable approach is to use a moment
measure, such as the mean $\langle l \rangle \equiv\sum_{l}lf_{l}$, and examine its
behavior as a function of expansion center location $\mathbf{a}$. It
is clear that as $\mathbf{a}$ moves very far away from the region
where most of the mass resides, the mass distribution acquires very
small angular scales, and the moment measure must increase without
bound. The moment measure is also obviously bounded below by 0. We
require that the choice of the expansion center location $\mathbf{a}$
should result in a value of that moment that is as small as possible.

From the point of view of practical computation, it turns out that
the most convenient moment for this purpose is
\begin{eqnarray}
\mu(\mathbf{a}) & \equiv & \langle l(l+1) \rangle (\mathbf{a})\nonumber \\
 & = & \sum_{l=0}^{\infty}l(l+1)f_{l}(\mathbf{a}).\label{eq:Compactness_Measure}
\end{eqnarray}
In order to find the ideal expansion
origin, we seek to minimize this ``spectral
compactness parameter'' with respect to expansion origin,
$\mathbf{a}$.  In the Appendix we show that the location that minimizes $\mu(\mathbf{a})$ is
approximately
\begin{eqnarray}
  \mathbf{a} & \approx & \frac{\int d^{3}\mathbf{x}\,\mathbf{x}\rho(\mathbf{x})^{2}}{\int d^{3}\mathbf{x}\,
    \rho(\mathbf{x})^{2}}\nonumber \\
  & \equiv & \left\langle \mathbf{x}\right\rangle
  _{\rho^{2}}. \label{eq:a_approx}
\end{eqnarray}

It is clear that this ``square-density weighted mean location'' is more biased
towards large condensations of mass than the ordinary CoM. It is
instructive to consider a simple example to illustrate the behavior of
$\left\langle \mathbf{x}\right\rangle _{\rho^{2}}$.  We imagine a
cubic box of side $L$, centered at a location $\mathbf{x}_{D}$ and
containing a uniform diffuse density $\rho_{D}$ corresponding to a
diffuse mass $M_{D}=L^{3}\rho_{D}$. The box also contains a sphere of
condensed mass of radius $r\ll L$ and uniform density $\rho_{C}$ (and
hence of mass $M_{C}=\frac{4\pi r^{3}}{3}\rho_{C}$) centered at a
location $\mathbf{x}_{C}$. It is straightforward to show that with
this mass configuration, the square-density-weighted CoM is
\begin{equation}
  \left\langle \mathbf{x}\right\rangle _{\rho^{2}}=\frac{M_{C}\left(\rho_{C}+2\rho_{D}\right)\mathbf{x}_{C}+M_{D}\rho_{D}
    \mathbf{x}_{D}}{M_{C}\left(\rho_{C}+2\rho_{D}\right)+M_{D}\rho_{D}}.\label{eq:Example_SDWCOM_1}
\end{equation}

If, for example, we assume the situation that prevails in CCSN simulations ---
that is, $M_{C}\sim M_{D}$, $\rho_{C}\gg\rho_{D}$, then this expression
becomes
\begin{equation}
  \left\langle \mathbf{x}\right\rangle _{\rho^{2}}=\mathbf{x}_{C}+\frac{\rho_{D}}{\rho_{C}}\frac{M_{D}}{M_{C}}
  \left(\mathbf{x}_{D}-\mathbf{x}_{C}\right)+\mathcal{O}\left(\left[\frac{\rho_{D}}{\rho_{C}}\right]^{2}\right).
  \label{eq:Example_SDWCOM_2}
\end{equation}
We can see that when the density contrast between $\rho_{D}$ and
$\rho_{c}$ is of many orders of magnitude, the square-density-weighted
CoM basically takes up residence at the center of the condensation.
This is the reason that the peak-density prescription for the
expansion center is so effective. By contrast, the usual CoM location
is the mass-weighted average of $\mathbf{x}_{D}$ and $\mathbf{x}_{C}$,
which can be well-separated from $\mathbf{x}_{C}$ if $M_{D}\sim
M_{C}$. Any such separation can obviously lead to troublesome angular
power at high values of $l$.

\section{Implementation of Multipole Poisson Solver in FLASH}
\label{sec:implement}

We use the FLASH hydrodynamic simulation framework \citep{Dubey:2009hh} to exhibit
the effects of the self-potential correction and the expansion centering schemes
described above.  In this section, we outline the implementation of the multipole gravity
solver in FLASH.  A more complete technical description of the algorithm is supplied in
the FLASH User's Guide\footnote{\texttt http://flash.uchicago.edu/site/flashcode/user\_support/}.

The discretized potential computation expressed in Eq.~(\ref{eq:Multipole_Discretized}) may be
separated into two distinct computations: the computation of an array of moments, and the
computation of the potential itself using the moments.  For notational convenience, we
introduce the solid harmonic functions
\begin{eqnarray}
R_{l m}(\mathbf{x}) & = & \sqrt{{4\pi\over {2l+1}}}r^l Y_{l m}({\bf n})
\label{eq:SolidHarmonicFunctionR} \\
I_{l m}(\mathbf{x}) & = & \sqrt{{4\pi\over {2l+1}}}{Y_{l m}({\bf n})\over r^{l+1}}.
\label{eq:SolidHarmonicFunctionI}
\end{eqnarray}

We will define multipole moments using a grid of concentric spheres
of increasing radii $r_\mu$, $\mu=1,2,\ldots$.  These radii are chosen at runtime 
depending on the nature of the mass distribution, and are not necessarily uniformly
spaced.  The spacing between radii is always more than one grid cell width, so that
the shells between successive spheres encompass multiple spherical layers of cells.
Given this grid, we may define the ``inner'' and ``outer'' multipole moment functions
\begin{eqnarray}
M^R_{l m}(r_\mu) & = & \sum_{\left|\mathbf{x}_{q^\prime}\right|\leq r_\mu} R_{l m}(\mathbf{x}_{q^\prime})m(q^\prime) \label{eq:Moment_Definition1} \\
M^I_{l m}(r_\mu) & = & \sum_{\left|\mathbf{x}_{q^\prime}\right|> r_\mu} I_{l m}(\mathbf{x}_{q^\prime})m(q^\prime), \label{eq:Moment_Definition2}
\end{eqnarray}
where $m(q^\prime)=\Delta_{q^\prime}^3\rho(\mathbf{x}_{q^\prime})$ is the mass of the
cell indexed by $q^\prime$.  

We further define $\mu^+(r)$ as the index $\mu$ of the  smallest of the $r_\mu$
exceeding $r$, and $\mu^-(r)$ as the index $\mu$ of the largest of the $r_\mu$ not exceeding
$r$, so that $\mu^+(r)-\mu^-(r)=1$.  We may then linearly interpolate the multipole moments:
\begin{eqnarray}
\widetilde{M}^{R,I}_{l m}(r)&\equiv& 
\frac{r-r_{\mu^-(r)}}{r_{\mu^+(r)}-r_{\mu^-(r)}}M^{R,I}_{l m}(r_{\mu^+(r)})+\nonumber\\
&&\frac{r-r_{\mu^+(r)}}{r_{\mu^-(r)}-r_{\mu^+(r)}}M^{R,I}_{l m}(r_{\mu^-(r)})\label{eq:Interpolate_Moments}
\end{eqnarray}

Using the interpolated moments, we write the discretized potential as
\begin{eqnarray}
\Phi(\mathbf{x}_q) & = & -G~{\mathcal Re}\left[\sum_{l m}\widetilde{M}^R_{l m}\left(|\mathbf{x}_q|)\right)I_{l m}^*(\mathbf{x}_q) \right. \nonumber \\
& + & \left.\sum_{l m}\widetilde{M}^{I*}_{l m}\left(|\mathbf{x}_q|)\right)R_{l m}(\mathbf{x}_q)\right].
\label{eq:Poisson_DiscreteSplit2}
\end{eqnarray}

The potential evaluation strategy is to first compute the multipole moments
from Eqs.~(\ref{eq:Moment_Definition1}) and (\ref{eq:Moment_Definition2}) using the chosen grid
of concentric spheres of radii $r_\mu$; then, at the second stage, use this array of moments to
compute the potential using Eq.~(\ref{eq:Poisson_DiscreteSplit2}).

The FLASH implementation of this strategy relies on explicitly real (sine and cosine)
versions of these formulae, which are described in the FLASH User's Guide.  The
real solid harmonic functions that arise are computed by recurrence relations that
follow from the Legendre function recurrence relations \citep{Arfken:2005uf}.  The
radial arguments of the solid harmonic functions are carefully scaled before the
recursion relations are applied, to prevent over- and underflows in large, highly-resolved
domains.

The implementation allows for different choices of spacing functions for the sphere
radii in different radial zones, so that, for example, the spacing could be linear
in an inner zone and logarithmic in an outer zone.  The range of possible choices is
described in the FLASH User's Guide.

As discussed in \S\ref{subsec:SelfPotError}, the potential evaluation 
described by Eq.~(\ref{eq:Poisson_DiscreteSplit2}) is always carried out at 
cell faces.  The cell-centered potential is then computed by averaging the 
potential of the faces bounding a cell.
Again, for multipole gravity algorithms based in spherical coordinates that compute the cell-centered potentials analytically \citep{Muller:1995jr}, rather than discretely, the self-potential error mitigated by our staggered computation approach should not be an issue.

The center of the multipole expansion is chosen by the FLASH solver to be the cell corner
nearest the square-density-weighted mean position, Eq.~(\ref{eq:a_approx}).  This choice
minimizes the spectral compactness, as described in \S\ref{subsection:Compactness}, and
also prevents any problematic potential evaluations at zero radius.

\section{Static Potential Test: MacLaurin Spheroids}
\label{sec:SimulationMacLaurin}

The analytic form of the gravitational potential of a stable,
rotationally symmetric, hydrostatic, uniform-density spheroid is due
to MacLaurin \citep[see][p. 77-]{Chandrasekhar:1987el}.  Such ``MacLaurin''
spheroids are useful for the validation of self-gravity solvers as
they provide an exact analytic solution against which to compare the
approximate calculated potentials. Here we consider the accuracy of
the multipole gravity solver for static MacLaurin spheroids.  We
compare the accuracy of the method using cell-centered potential
solves to that of using face-centered solves.


The exact gravitational potential for a point within a MacLaurin spheroid of density $\rho$ is:
\begin{eqnarray}
  \Phi({\bf x}) = \pi G \rho [& & 2A_1 a_1^2 - A_1(x^2+y^2) \nonumber \\
        &+& A_3(a_3^2-z^2)],
\end{eqnarray}
where $a_1$, $a_2$, and $a_3$ are the semi-major axes of the spheroid
and $a_1 = a_2 > a_3$. Here
\begin{eqnarray}
  A_1 &=& \frac{\sqrt{1-e^2}}{e^3} \sin^{-1}e - \frac{1-e^2}{e^2}\ , \\
  A_3 &=& \frac{2}{e^2} - \frac{2\sqrt{1-e^2}}{e^3} \sin^{-1}e\ ,
\end{eqnarray}
where $e$ is the ellipticity of a spheroid:
\begin{equation}
  e = \sqrt{1 - \left( \frac{a_3}{a_1} \right) ^2}\ .
\end{equation}
For a point outside the spheroid, potential is: 
\begin{equation}
  \begin{split}
    \Phi({\bf x}) &= \frac{2a_3}{e^2} \pi G \rho \left[a_1 e
      \tan^{-1}h     \vphantom{\frac{1}{1}} \right. \\
    &- \frac{1}{2} ( (x^2+y^2)(
      \tan^{-1}h - \frac{h}{1+h^2} )  \\
     &+ 2z^2 (h-\tan^{-1}h) ) ]\ ,
    \end{split}
\end{equation}
where 
\begin{equation}
  h = \frac{a_1 e}{\sqrt{a_3^2 + \lambda}}\ ,
\end{equation}
and $\lambda$ is the positive root of the equation
\begin{equation}
  \frac{x^2}{a_1^2 + \lambda} + \frac{y^2}{a_2^2 + \lambda} + \frac{z^2}{a_3^2 + \lambda} = 1\ .
\end{equation}


\begin{figure}
  \centering
  \includegraphics[width=3.45in]{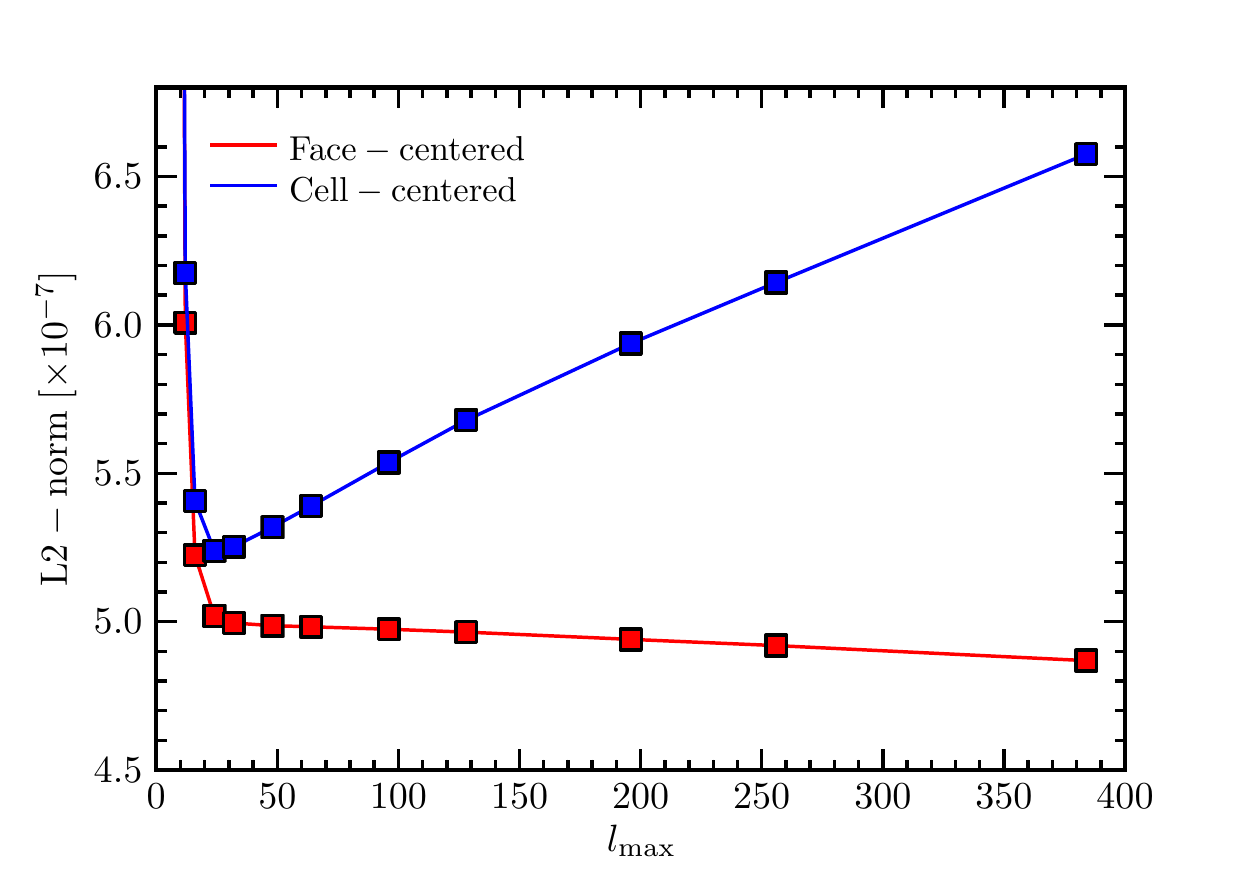}
  \caption{L2-norm error for the 2D MacLaurin spheroid problem with
    $e=0.9$.  The blue line and boxes are for potential solvers at
    cell centers, the red line and boxes are for face-centered
    potential solves.  The expected approximate linear growth in the
    error due to the potential self-energy
    [c.f. Eq. (\ref{eq:Self_Energy_Divergence})].  This growth in the
    error is absent for face-centered potential calculations and the
    error continues to decrease with $l_{\rm max}$.  Note also that
    the magnitude of the L2-norm error is smaller for the
    face-centered calculation at every $l_{\rm max}$.}
  \label{fig:macLaurin}
\end{figure}

For the present tests we consider a spheroid of uniform density $\rho
= 1$ g cm$^{-3}$ embedded in a background of vanishing density, $\rho
\approx 0$.  We use an eccentricity 0.9 in 2D cylindrical
geometry and compare the L2-norm error of the cell-centered potential
calculation with that of the face-centered potential calculation.
Figure \ref{fig:macLaurin} shows the results. These tests span a very
large range in \lmax, from 0 to 384.  We find that at every value of
\lmax\ the face-centered calculation yields a smaller L2-norm error,
i.e., it is more accurate.  And at high values of \lmax, beyond about
24, the cell-centered calculation error {\it increases}
with higher \lmax.  The character of this increase is very nearly
linear, just as we would expect based on equation
(\ref{eq:Self_Energy_Divergence}). The face-centered calculation, on
the other hand, results in an error that continues to decrease with
\lmax, i.e., the accuracy of the calculation converges with \lmax.

\begin{figure*}
  \centering
  \begin{tabular}{cc}
    \includegraphics[width=3.45in,trim = 1.in 1.5in 1in .75in, clip]{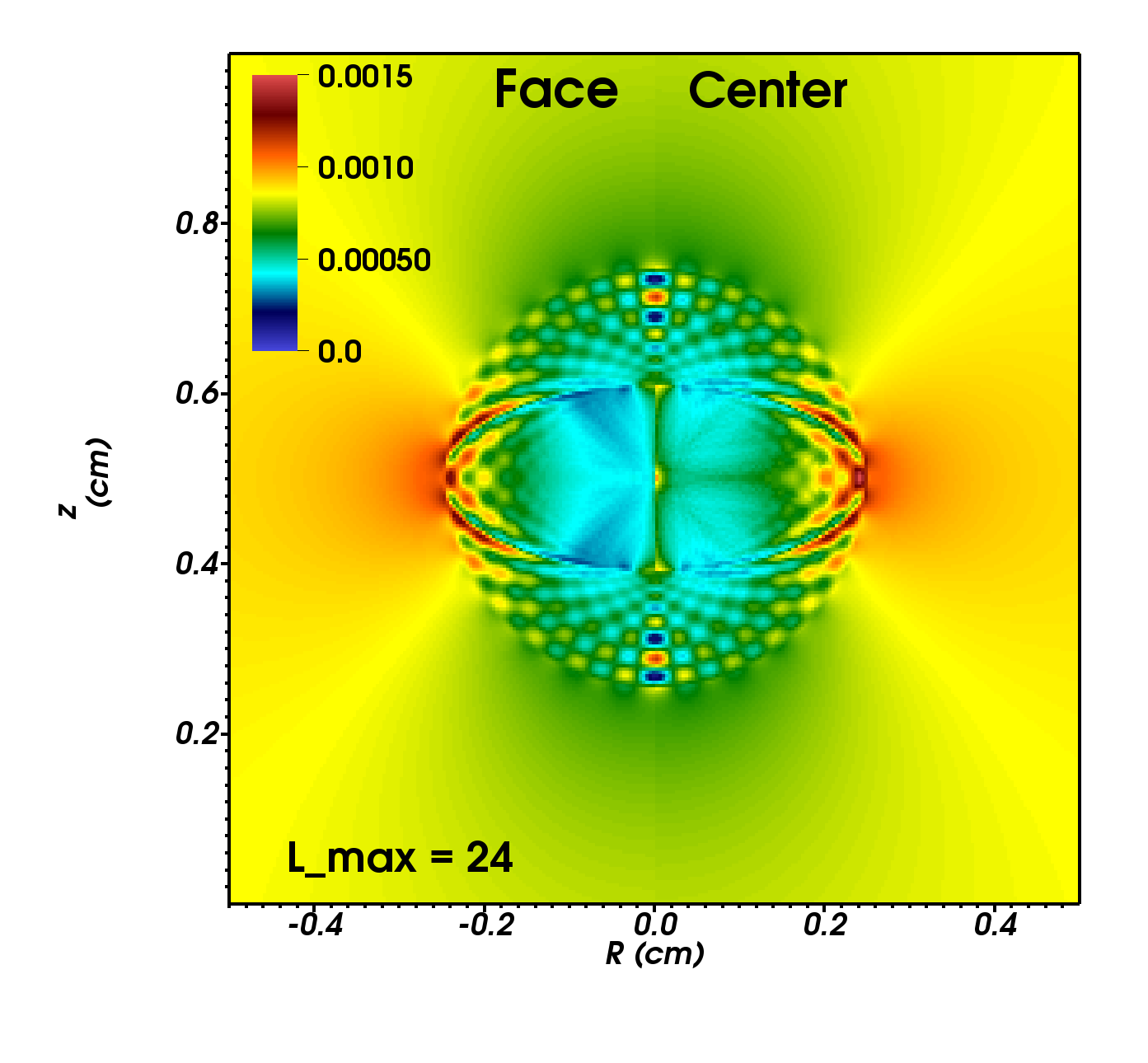}
    \includegraphics[width=3.45in,trim = 1in 1.5in 1in .75in, clip]{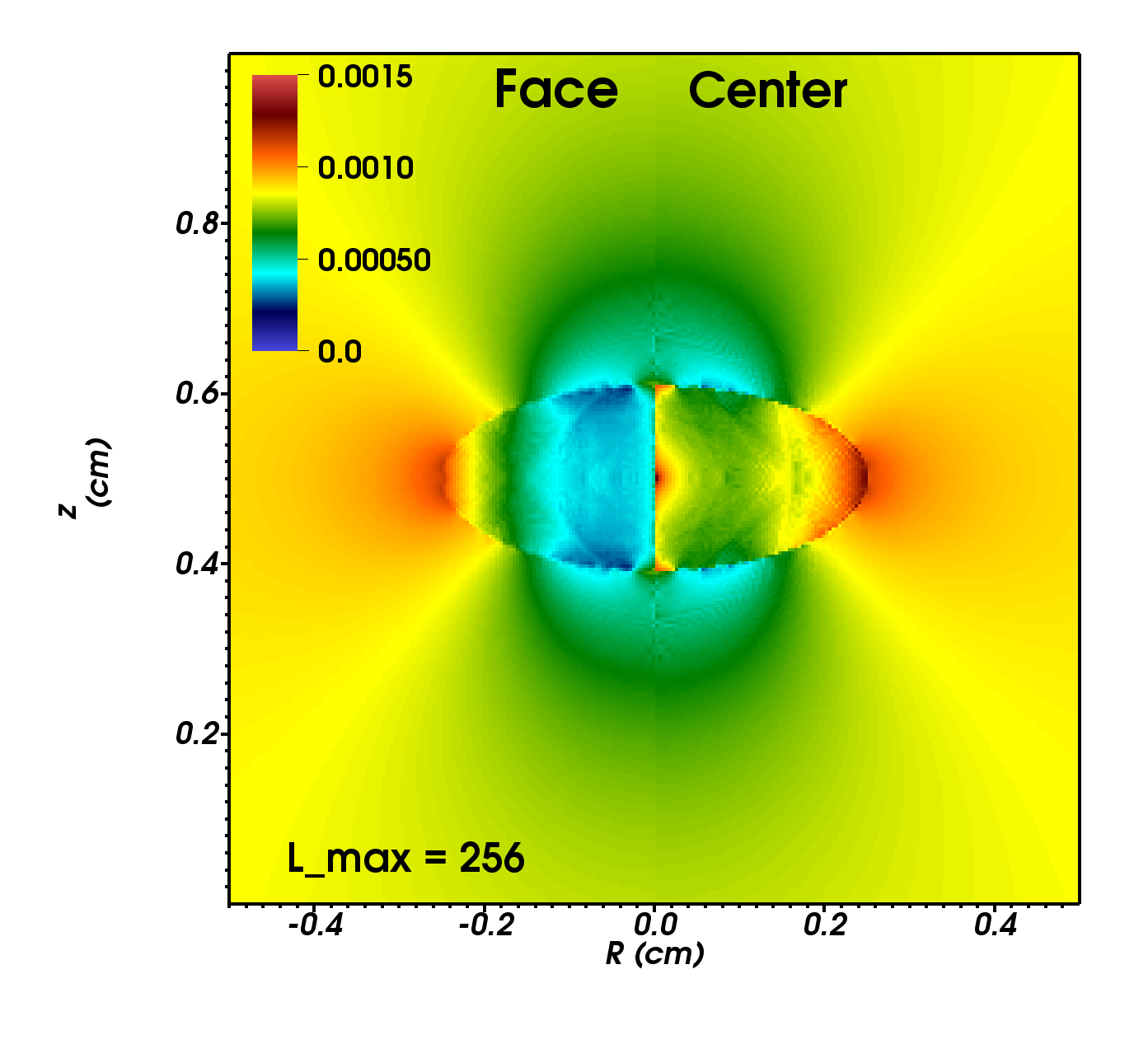}
  \end{tabular}
  \caption{Comparison of normalized errors in the gravitational
    potential for a MacLaurin spheroid of eccentricity $e=0.9$.  The
    left panel compares the error for face-center potential evaluation
    (left) to cell-centered evaluation (right) for \lmax = 24.  In the
    right panel we show the same comparison but for \lmax = 256.
    Using face-centered potential calculations results in reduction in
    both the peak error and the L2-norm error.  Using cell-centered
    calculation results in an error near the center of the multipole
    expansion that grows with \lmax, just as we would predict based on
    equation (\ref{eq:Self_Energy_Divergence}).}
  \label{fig:macLaurin2D}
\end{figure*}

Further evidence that the self-potential error isolated and exhibited in
equation (\ref{eq:Self_Energy_Divergence}) is real and present in the
cell-centered potential calculation is given by inspection of the
normalized error in the potential.  In Figure \ref{fig:macLaurin2D} we
show pseudocolor plots of the normalized error in the potential for a
MacLaurin spheroid with $e = 0.9$ for two different values of \lmax,
and compare cell-centered and face-centered potential calculations.
The self-potential error of equation (\ref{eq:Self_Energy_Divergence})
predicts that the largest errors occur near the center of
the multipole expansion.  In the case of the 2D cylindrical spheroid
of Figure \ref{fig:macLaurin2D} this is $R=0$, $z=0.5$.  We see that
this is precisely the case.  For the cell-centered calculation there
is a large normalized error at the center of the spheroid that is {\it
  absent} in the face-centered calculation.  Additionally we see that
the magnitude of this error {\it increases} for larger \lmax\ in the
cell-centered case.


\section{Dynamic Potentials: Core-Collapse Supernovae}
\label{sec:ccsn}

Static potentials for which analytic solutions are known are useful in
verifying the accuracy of the self-gravity solver but we also seek to
test if our novel handling for the errors present in multipole
approximations have a positive impact on dynamical simulations that
hinge critically on self-gravity.  For this we turn to CCSN
simulations.  Having established in Section
\ref{sec:SimulationMacLaurin} that face-centered potential
calculations avoid the self-potential error, resulting in greater
accuracy of the potential and convergence with increasing \lmax, we
focus only on the face-centered potential calculation approach for the
CCSN simulations.  We test the impact of different multipole expansion
centering on the CCSN problem by running simulations with different
values of \lmax\ for three different expansion centers: the center of
mass, fixed at the coordinate origin, and the square-density-weighted
mean location (SDML).

In our finite-volume Eulerian approach, gravity is coupled to the
hydrodynamic calculation via source terms on the right-hand-sides of
the momenta and energy equations.  In FLASH, these source terms are
included in the Riemann solver as corrections to the intermediate cell
face states that are used in calculating time-centered face fluxes of
conserved quantities.  We have modified the coupling of gravity and
hydro in FLASH in the following way.  Previous versions of FLASH {\it
  extrapolated} the gravitational acceleration to the time step
midpoint ($n+1/2$) using the current ($n$) and previous ($n-1$) time
step accelerations.  This approach is formally only first-order
accurate in time.  We have adopted instead the second-order accurate
approach of {\it interpolating} the acceleration to the time step
midpoint by first updating the density field via the continuity
equation, then reevaluating the gravitational potential, then
finishing the finite-volume update of momenta and energy with
time-centered gravitational accelerations interpolated to $n+1/2$
using the $n$ and $n+1$ state accelerations.  This is the approach
used in, e.g., CASTRO \citep{Almgren:2010du}.  Since this approach
still utilizes source terms, the scheme is not expected to conserve
momenta and energy perfectly.  Such conservation can be achieved by
using the method of, e.g., \citet{Jiang:2013ii}.

For these simulations we use the approach of \citet{Couch:2013df,
  Couch:2012un}.  We follow the evolution from the collapse phase
through core bounce and into shock revival by neutrino heating.  We
assume simple local neutrino heating/cooling as introduced by
\citet{Murphy:2008ij} with an exponential cutoff of the neutrino
source terms at high density.  Deleptonization is accounted for using
the density-dependent parameterization of \citet{Liebendorfer:2005ft},
both pre- and post-bounce.  The only modification we make to the
method of \citet{Couch:2013df, Couch:2012un} is to weight the
density-dependent neutrino source term cutoff so that we achieve a
critical luminosity for explosion closer to that of
\citet{Murphy:2008ij}, as was also done in \citet{Hanke:2012dx}.  All
of our simulations are carried out in 2D cylindrical geometry with a
maximum resolution of 0.5 km and we use the 15 \Msun\ progenitor of
\citet{Woosley:1995jn}. We use a fixed neutrino luminosity of
$2.2\times 10^{52}$ erg s$^{-1}$.  

\begin{figure*}
\centering
\includegraphics[width=7.5in,trim= .25in .15in .25in .25in, clip]{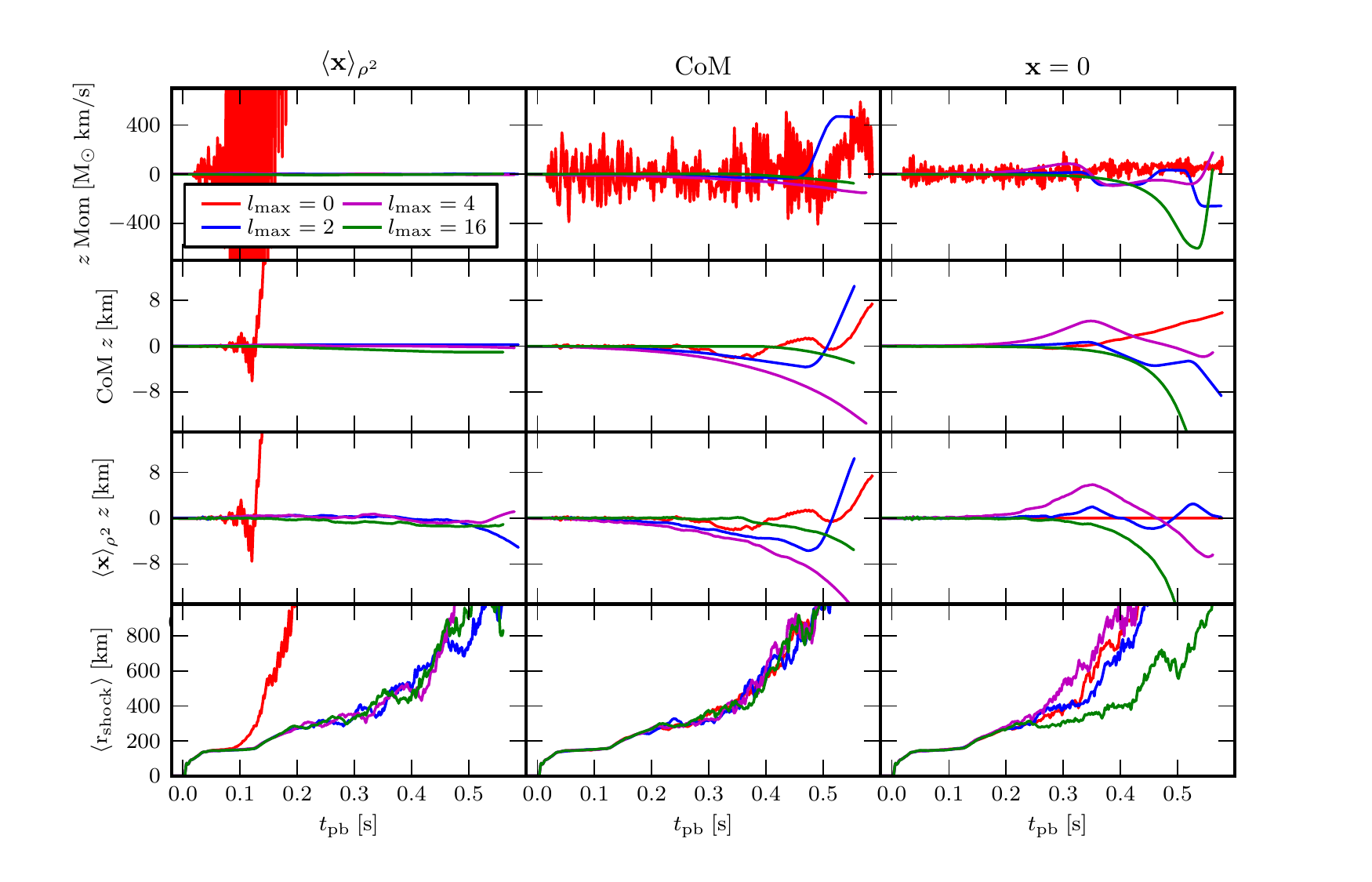}
\caption{Various simulation diagnostics for the CCSN simulations as
  functions of post-bounce time for the three multipole expansion
  centering approaches we test.  The columns represent
multipole expansion centering on the SDML (left), CoM (middle), and
coordinate origin (right).  The rows show, from top to botton,
$z$-momentum, CoM $z$ coordinate, $\langle \mathbf{x}
\rangle_{\rho^2}$ $z$ coordinate, and the average shock radii. }
\label{fig:expansionOrigin}
\end{figure*}

In Figure \ref{fig:expansionOrigin} we graphically present the results
of the CCSN simulations for several values of \lmax\ and multiple
expansion origins.  We show as function of post-bounce time the
$z$-coordinate of both the $\langle \mathbf{x} \rangle_{\rho^2}$ and
the CoM along with the total $z$-momentum and average shock radius.
For these 2D axisymmetric calculations initialized from
spherically-symmetric initial conditions the CoM should remain fixed
at the coordinate origin, which is simply a restatement of the
conservation of total $z$-momentum.  We find that for \lmax $> 0$
centering the multipole expansion on the square-density-weighted mean
location results in {\it dramatically} improved conservation of
$z$-momentum.  For other choices of expansion center the $z$-momentum
non-conservation can be in excess of 400 \Msun km s$^{-1}$.  This
spurious momentum is about the same as what is observed for typical
neutron stars!  The magnitude of the momentum non-conservation is
indiscernible for the case of centering on $\langle \mathbf{x}
\rangle_{\rho^2}$, though conservation is not perfect as reflected by
the slight drift in the CoM.

These simulations result in non-symmetric explosions and so we 
expect that the PNS will receive a kick. The $\langle \mathbf{x}
\rangle_{\rho^2}$ tracks very well the center of the PNS and so its
motion can be regarded as that of the PNS.  Much larger kicks are
imparted to the PNS for the CoM and $\mathbf{x}=0$ cases, for which we
measure large non-conservations of momenta.  The PNS also begins its
motion much earlier than the SDML case.  The kick of the PNS is
obviously affected by the momentum non-conservation.  It is worth
noting that for CoM centering, the conservation of momentum improves
with increasing \lmax, but even for \lmax $= 16$ the CoM still moves by
about 3 km, or 6 numerical zones while the CoM barely moves by one
zone for any \lmax\ for SDML centering.

The SDML centering is obviously superior to other centering choices
for \lmax~$> 0$, but equally obvious is its utter failure for \lmax =
0. In the case of monopole gravity centering the expansion on SDML
allows the PNS to move too easily away from the CoM while not
correctly accounting for the strong dipole term that would result and
pull the PNS back.  We also see that for centering at the coordinate
origin and \lmax = 0, the PNS is held fixed in place and does not
receive a kick.  The CoM still moves in this case, reflecting
non-conservation of momentum.  Centering on the coordinate origin also
displays {\it divergent} behavior with increasing \lmax: higher values
result in greater non-conservation of momentum and greater spurious
motion of the PNS.

The average shock radius histories for coordinate origin centering
are also highly variable with respect to changes in \lmax.  The other
expansion centering approaches yield highly consistant shock radius
histories for all values of \lmax, save for \lmax $= 0$ in the SDML
case.

\begin{figure}
\centering
\includegraphics[width=3.25in, trim= .05in .25in .33in .4in, clip]{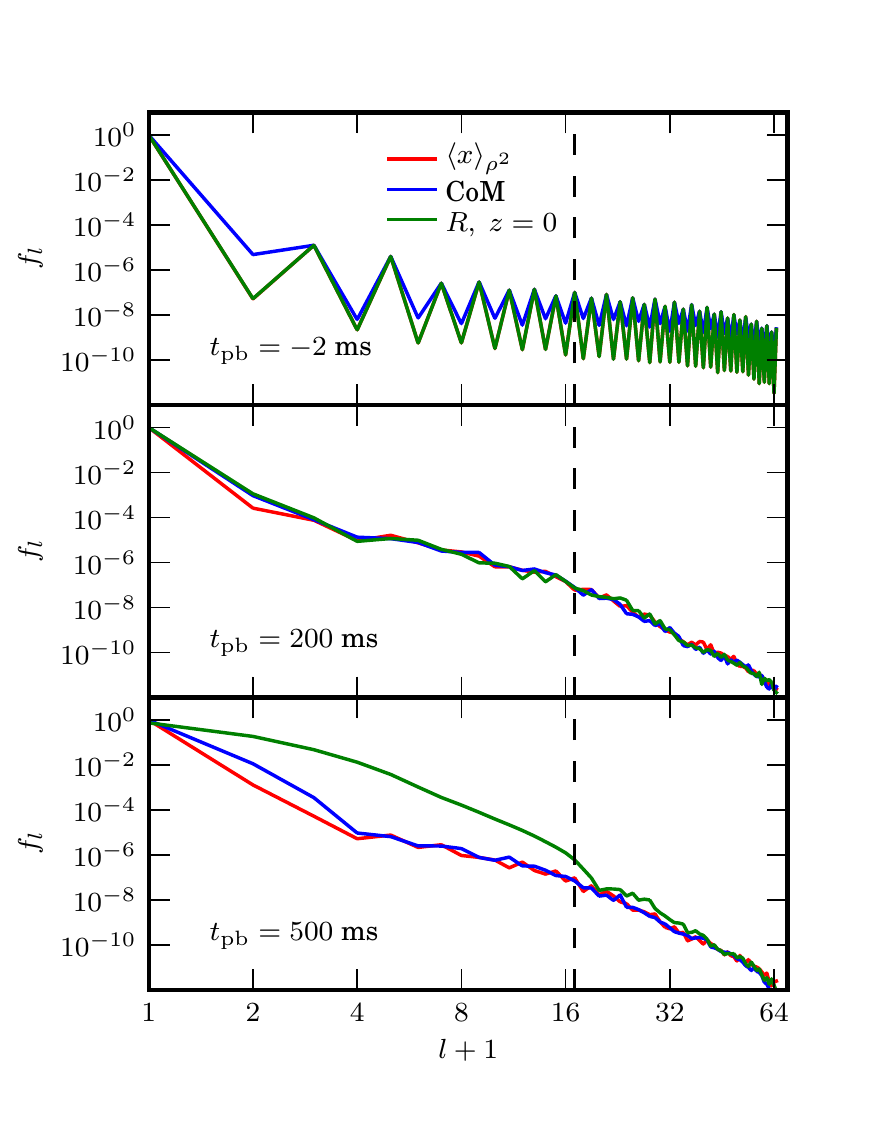}
\caption{Normalized potential energy spectra at three different times
  for the three different multipole expansion centering approaches. In
  the top panel, the red and green lines are indistinguishable. }
\label{fig:spectra}
\end{figure}

Our choice of the SDML for the multipole expansion centering is
motivated by our minimization of the spectral compactness, $\mu$,
introduced in Section \ref{sec:mpoleExp}.  This metric, defined as
$\langle l(l+1) \rangle$, measures the concentration of total
gravitational potential energy at low multipole
orders. Our analysis in Section \ref{sec:mpoleExp} indicates that
centering the expansion on the SDML should maximize the amount of
total potential energy from low orders, i.e., yield the most spherical
representation of the gravitational potential.  To test this for the
CCSN simulations we compute the normalized potential energy spectra,
$f_l \equiv \mathcal{E}_l / \mathcal{E}$, for the three
different expansion centering approaches at three different times,
shown in Figure \ref{fig:spectra}.  All of the simulations in Figure
\ref{fig:spectra} were run with \lmax = 16 but we compute the spectra
out to $l = 64$ and indicate $l = 16$ by the vertical dashed line.
Prior to core bounce ($t_{\rm pb} = -2$ ms) the spectra are highly
concentrated at $l = 0$, as expected for the spherically-symmetry mass
distribution, and the odd multipoles are much reduced due to the
symmetry. By 200 ms post-bounce the spectra remain very similar except
for the slightly reduced power in $l = 1$ and greater power in $l = 0$
for the SDML case.  The shock radii are also very similar at this time
(see Figure \ref{fig:expansionOrigin}).  The differences in the
centering approaches are more obvious at 500 ms after runaway shock
expansion has begun. The reduced power at multipoles
greater than 0 is evident for the SDML case while centering on the
coordinate origin results in a spreading of the spectrum to larger
multipoles.

\section{Conclusions}
\label{sec:conclusions}

We have identified and corrected  two sources of
error arising in general discretized multipole approximations to
Poisson's equation.  The first error results from assuming that all
the mass in a computational zone resides at the cell center and then
evaluating the potential at the same point.  Inspection of the Green's
function for the continuous Poisson equation makes obvious that this
error has its origin in the divergent $\lvert \mathbf{x}-\mathbf{x}'
\rvert^{-1}$ term.  This term is explicitly absent from the
discretized equations but the error it induces is still lurking in the
method.  We show that this error is proportional to the mass in a
zone, divided by the distance of the zone center from the origin of
the multipole expansion, multiplied by \lmax $+ 1$.  This error therefore
grows rather than shrinking as the number of
terms retained in the truncated expansion increases.  We show that the
``self-potential'' error can be corrected by evaluating the
gravitational potential at cell faces, where no mass has been located,
rather than cell centers.  The cell-centered potential is then found
by averaging the potential at the cell-bounding faces.  Using MacLaurin spheroids, for which exact analytic
potential solutions are known, we show that this approach improves the accuracy
of the potential calculation and leads to convergence of the solution
with increasing \lmax, i.e., the self-potential error is {\it
  eliminated}.  

The second error we identify has to do with a poor selection of the
multipole expansion origin.  By suggesting a useful metric, the
spectral compactness $\mu = \langle l(l+1) \rangle$, characterizing
the symmetry of the potential we find that the optimal location for
the origin that minimizes $\mu$ is the square-density-weighted mean
location, $\langle \mathbf{x} \rangle_{\rho^2}$.  For diffuse mass
distributions, or distributions in which the total mass in the computational
domain is dominated by a single condensation, this location is not too different from the center of mass, the
common choice for multipole expansion origin.  For high-mass
condensations embedded in high-mass diffuse flows, such as occur in CCSN simulations that include the
proto-neutron star, the $\langle \mathbf{x} \rangle_{\rho^2}$ is close to
the peak density of the high-mass condensate.  Using a series of CCSN
simulations we demonstrate the superiority of locating the expansion
center at the SDML: momentum conservation is {\it dramatically}
improved resulting in significantly different kicks imparted to the
PNS by the development of asymmetric explosions.  CCSN simulations
that include the PNS are especially susceptible the two errors we
discuss because of the enormous mass density in few zones near
the expansion origin. 

Our computational approach is embedded in an Eulerian hydrodynamic 
framework.  Nevertheless, the multipole approach to the solution of 
the Poisson equation is quite general, and its numerical 
implementation stands apart from the specific numerical hydrodynamic 
scheme employed here. It follows that the improvements we describe 
above to the discretized multipole approximation to Poisson's 
equation are generally applicable.  In particular, the optimal choice of
expansion center is relevant to {\it all} simulations that 
employ multipole approach for calculating self-gravity, and the face-centering
of the potential calculation is relevant to all such approaches that are
grid-based and do not evaluate potentials analytically, as can be done in spherical geometry \citep{Muller:1995jr}.  

It is important to note that the momentum non-conservation, and
concomitant erroneous motion of the PNS, is due to the movement of the
PNS away from the origin of the multipole expansion.  In spherical
geometry where the PNS is unable to move away from the origin, or in
CCSN simulations that excise the PNS, we do not expect to see such bad
momentum non-conservation.  We, therefore, do not expect previous
studies of PNS kicks that excise the PNS from
the domain in spherical geometry \citep{{Scheck:2004ez},{Scheck:2006iz}, Wongwathanarat:2010ch, {Wongwathanarat:2013fx}} to
suffer from the inaccuracies we here uncover and correct. Likewise our
results have no bearing on PNS kick studies that do not utilize
multipole gravity solvers \citep{{Nordhaus:2010kq},
  {Nordhaus:2012dx}}.

The multipole approach is appropriate for systems wherein the mass
distribution is approximately spherical, so that a spherical harmonic
expansion can be expected to reach high accuracy after a moderate
number of terms.  For such problems it has substantial benefits over
other approaches for solving Poisson's equation, such as multigrid or
tree methods, because it is comparatively inexpensive.  For the
time-dependent CCSN simulations described in Section \ref {sec:ccsn}
the multipole implementation we present in Section \ref
{sec:implement} requires less than 7\% of the time to calculate the
hydrodynamics.  More exact multigrid and tree methods can dominate the
computational expense of simulations utilizing them
\citep[c.f.][]{Ricker:2008eq}.  By incorporating the two essential
reforms of the multipole algorithm we present the method can deliver
on its promise of accurate calculation of self gravity while also
retaining its efficient computation.

\acknowledgements 
The authors thank Paul Ricker, Thomas Janka, and Annop Wongwathanarat for helpful
conversations.  We especially thank our referee, Ewald M\"uller, for a careful review that improved this article. SMC is supported by NASA through Hubble Fellowship
grant No. 51286.01 awarded by the Space Telescope Science Institute,
which is operated by the Association of Universities for Research in
Astronomy, Inc., for NASA, under contract NAS 5-26555.  This work was
supported in part by the National Science Foundation under grant
AST-0909132.  This work was supported in part at the University of
Chicago by the US Department of Energy (DOE) under contract B523820 to
the NNSA ASC/Alliances Center for Astrophysical Thermonuclear Flashes.
The software used in this work was in part developed by the DOE
NNSA-ASC OASCR Flash Center at the University of Chicago.  This
research used computational resources at ALCF at ANL, which is
supported by the Office of Science of the US Department of Energy
under Contract No. DE-AC02-06CH11357.

\appendix

\section{Gravitational Binding Energy as Angular Spectrum}

Our proposed angular decomposition of the global spectral of the gravitational
field is proportional to $\mathcal{E}_{l}$, where
\begin{eqnarray}
  \mathcal{E}_{l} & \equiv & \frac{G}{2}\frac{4\pi}{2l+1}\sum_{m=-l}^{l}\int d^{3}\mathbf{x}\, d^{3}\mathbf{x}^{\prime}\,
  \rho(\mathbf{x})\rho(\mathbf{x}^{\prime})\, Y_{lm}(\mathbf{n})Y_{lm}(\mathbf{n}^{\prime})^{*}g_{l}(r,r^{\prime}),
\label{eq:E_l_1_appendix}
\end{eqnarray}
and $g_{l}(r,r^{\prime})$ is the function given in Eq.~(\ref{eq:g_factor}).

In order for this choice of spectral decomposition of the field to give rise
to a sensible distribution,
it is necessary to establish that $\mathcal{E}_{l}\ge0$ for all $l$.
To do this, we write Eq.~(\ref{eq:E_l_1_appendix}) as
\begin{equation}
  \mathcal{E}_{l}=\frac{G}{2}\frac{4\pi}{2l+1}\sum_{m=-l}^{l}\int_{0}^{\infty}r^{2}dr\int_{0}^{\infty}r^{\prime2}dr^{\prime}
  g_{l}(r,r^{\prime})\,\eta_{lm}(r)\eta_{lm}(r^{\prime})^{*},\label{eq:E_L_2}
\end{equation}
where we have defined the moments
\begin{equation}
  \eta_{lm}(r)\equiv\oint d^{2}\mathbf{n}\,\rho(r\mathbf{n})Y_{lm}(\mathbf{n}).\label{eq:Moment_Integral}
\end{equation}
From Eq.~(\ref{eq:E_L_2}), we see that the requirement that
$\mathcal{E}_{l}\ge0$ is equivalent to requiring
positive-semi-definiteness of the integral operator $\hat{g}_{l}$,
whose action on a function $\eta(r)$ is
$\left[\hat{g}_{l}\circ\eta\right](r)=\int_{0}^{\infty}r^{\prime2}dr^{\prime}\,
g_{l}(r,r^{\prime})\eta(r^{\prime})$.  In other words, we must have
$(\eta,\hat{g}_{l}\circ\eta)=\int_{0}^{\infty}r^{2}dr\,\eta(r)\left[\hat{g}_{l}\circ\eta
\right](r)\ge0$.  But $(2l+1)^{-1}\hat{g}_{l}$ is the inverse of the
radially-separated Laplacian differential operator
$\mathcal{L}_{l}\equiv-\frac{1}{r^{2}}\frac{\partial}{\partial
  r}\left(r^{2}\frac{\partial}{\partial
    r}\right)+\frac{l(l+1)}{r^{2}}$, for which $g_{l}(r,r^{\prime})$
is the Green's function: $\left[\mathcal{L}_{l}\circ\hat{g}_{l}\right]
(r,r^{\prime})=\frac{2l+1}{r^2}\delta(r-r^{\prime})$.  Furthermore, we may
easily show that $\mathcal{L}_{l}$ is positive-definite,
$\left(\eta,\mathcal{L}_{l}\circ\eta\right)>0$,
for the boundary conditions of interest here (finite at the origin,
zero at infinity) by means of an integration by parts. Setting
$\hat{g}_{l}\circ\eta\equiv(2l+1)\chi$, so that
$\eta=\mathcal{L}_{l}\circ\chi$ we therefore
have
\begin{equation}
  (\eta,\hat{g}_{l}\circ\eta)=(2l+1)\times\left(\mathcal{L}_{l}\circ\chi,\chi
  \right)>0.\label{eq:Ll_positive_definite}
\end{equation}
Since $\hat{g}_{l}$ is a positive-definite integral operator, it
follows immediately from Eq.~(\ref{eq:E_L_2}) that $\mathcal{E}_{l}>0$
for all $l$.

The normalized distribution over $l$ $f_{l}\equiv\mathcal{E}_{l}/\mathcal{E}$
is therefore a sensible measure of the angular spectrum in a gravitating
mass distribution. The total
binding energy $\mathcal{E}$ is obviously independent of the expansion
center position $\mathbf{a}$. The individual terms $\mathcal{E}_{l}$
in the decomposition are certainly functions of $\mathbf{a}$, however,
so that the spectral distribution is also dependent on $\mathbf{a}$.
We will therefore write this dependence as $\mathcal{E}_{l}(\mathbf{a})$
explicitly below.

When calculating the spectrum $f_l$ empirically from a mass
distribution, as we do in \S\ref{sec:ccsn}, there is a subtle source
of error to be guarded against, which is traceable to discretization
noise. The effect comes about because, as remarked earlier, the mass
of each cell, which represents a volume integral of some smooth,
nearly constant mass density function over the cell, is represented in
the numeric quadratures of the multipole algorithm as a
Dirac-$\delta$-function at the cell center. Obviously, an
infinitely-narrow density peak is capable of contributing power to
arbitrarily-high multipole orders $l$, whereas the cell's contribution
to the angular spectrum due to the underlying, nearly constant density
function should cut off rapidly above some angular scale. The error
therefore manifests itself in the spectrum $f_l$ as a noisy positive
DC-offset level at high $l$-values.  In order to exhibit normalizable
spectra, it is necessary to remove this error.  This can be done by
observing that a cell with index $q$, of size $\Delta_q$, located at a
distance $r_q$ from the center of the expansion, subtends an angle
$\theta_q\sim\Delta_q/r_q$ at the center.  We should not expect such a
cell to contribute anything but noise to multipoles of order $l >
2\pi/\theta_q$.  Discarding such terms from the multipole moment
contribution of these cells, the DC offset is removed, and
normalizable spectra such as the ones shown in \S\ref{sec:ccsn} are
recovered.

\section{Extremizing Spectral Compactness}

As asserted in \S\ref{subsection:Compactness}, from the point of 
view of practical computation, it turns out that the most convenient 
moment for the purpose of quantifying angular spectrum compactness is
\begin{eqnarray}
\mu(\mathbf{a}) & \equiv & <l(l+1)>(\mathbf{a})\nonumber \\
 & = & \sum_{l=0}^{\infty}l(l+1)f_{l}(\mathbf{a}).\label{eq:Compactness_Measure_appendix}
\end{eqnarray}
The reason this is convenient is because the term $l(l+1)$ arises naturally from
the application of the Laplacian to the spherical harmonic expansion
of the Green's function $\left|\mathbf{x}-\mathbf{x}^{\prime}\right|^{-1}$:
\begin{eqnarray*}
  -4\pi\delta^{3}\left(\mathbf{x}-\mathbf{x}^{\prime}\right) & = & \nabla^{2}\left|\mathbf{x}-\mathbf{x}^{\prime}\right|
  ^{-1}\\
  & = & \nabla^{2}\left\{ \sum_{l=0}^{\infty}\sum_{m=-l}^{l}\frac{4\pi}{2l+1}Y_{lm}(\mathbf{n})Y_{lm}(\mathbf{n}
    ^{\prime})^{*}g_{l}(r,r^{\prime})\right\} \\
  & = & \sum_{l=0}^{\infty}\sum_{m=-l}^{l}\frac{4\pi}{2l+1}\left[\frac{1}{r^{2}}\frac{\partial}{\partial r}
    \left(r^{2}\frac{\partial}{\partial r}\right)-\frac{l(l+1)}{r^{2}}\right]Y_{lm}(\mathbf{n})Y_{lm}(\mathbf{n}^{\prime})^{*}
  g_{l}(r,r^{\prime})\\
  & = & \frac{1}{r^{2}}\frac{\partial}{\partial r}\left(r^{2}\frac{\partial}{\partial r}\right)\left|\mathbf{x}-\mathbf{x}
    ^{\prime}\right|^{-1}-\frac{1}{r^{2}}\sum_{l=0}^{\infty}\sum_{m=-l}^{l}\frac{4\pi}{2l+1}l(l+1)Y_{lm}(\mathbf{n})Y_{lm}
  (\mathbf{n}^{\prime})^{*}g_{l}(r,r^{\prime})
\end{eqnarray*}
so that
\begin{eqnarray}
  \sum_{l=0}^{\infty}\sum_{m=-l}^{l}\frac{4\pi}{2l+1}l(l+1)Y_{lm}(\mathbf{n})Y_{lm}(\mathbf{n}^{\prime})^{*}g_{l}
  (r,r^{\prime}) & = & \frac{\partial}{\partial r}\left(r^{2}\frac{\partial}{\partial r}\right)\left|\mathbf{x}-\mathbf{x}
    ^{\prime}\right|^{-1}+4\pi r^{2}\delta^{3}\left(\mathbf{x}-\mathbf{x}^{\prime}\right).\nonumber \\
\label{eq:trick}
\end{eqnarray}
Combining Eqs.~(\ref{eq:E_l_1_appendix}) and (\ref{eq:trick}) with Eq.~(
\ref{eq:Compactness_Measure_appendix}), and setting $\mathbf{a}=\mathbf{0}$
temporarily, we obtain
\begin{equation}
  \mu(\mathbf{0})=\frac{G}{2\mathcal{E}}\int d^{3}\mathbf{x}\, d^{3}\mathbf{x}^{\prime}\,\rho(\mathbf{x})\rho(\mathbf{x}
  ^{\prime})\left\{ 4\pi r^{2}\delta^{3}\left(\mathbf{x}-\mathbf{x}^{\prime}\right)+\frac{\partial}{\partial r}
    \left(r^{2}\frac{\partial}{\partial r}\right)\left|\mathbf{x}-\mathbf{x}^{\prime}\right|^{-1}\right\} .\label{eq:mu_0_1}
\end{equation}
At the cost of some algebra, we may evaluate the second term in Eq.~(\ref{eq:mu_0_1}).
We obtain
\begin{equation}
  \mu(\mathbf{0})=\frac{G}{2\mathcal{E}}\int d^{3}\mathbf{x}\, d^{3}\mathbf{x}^{\prime}\,\rho(\mathbf{x})\rho(\mathbf{x}
  ^{\prime})\left\{ 4\pi r^{2}\delta^{3}\left(\mathbf{x}-\mathbf{x}^{\prime}\right)+\frac{-(\mathbf{x}\cdot\mathbf{x}
      ^{\prime})^{2}-3r^{2}r^{\prime^{2}}+2(r^{2}+r^{\prime2})\mathbf{x}\cdot\mathbf{x}^{\prime}}{\left|\mathbf{x}-\mathbf{x}
        ^{\prime}\right|^{5}}\right\} .\label{eq:mu_0_2}
\end{equation}

After making the substitution
$\mathbf{x}^{\prime}=(\mathbf{x^{\prime}}-\mathbf{x})+\mathbf{x}$ in
the numerator of the second term, some further algebra
yields
\begin{eqnarray}
  \mu(\mathbf{0}) & = & \frac{G}{2\mathcal{E}}\int d^{3}\mathbf{x}\, d^{3}\mathbf{x}^{\prime}\,\rho(\mathbf{x})
  \rho(\mathbf{x}^{\prime})\Biggl\{4\pi r^{2}\delta^{3}\left(\mathbf{x}-\mathbf{x}^{\prime}\right)+\sum_{i,k=1}^{3}x_{i}
  x_{k}\frac{3(x_{i}^{\prime}-x_{i})(x_{k}^{\prime}-x_{k})-\left|\mathbf{x}-\mathbf{x}^{\prime}\right|^{2}\delta_{ik}}
  {\left|\mathbf{x}-\mathbf{x}^{\prime}\right|^{5}}\nonumber \\
  &  & \hspace{3.5cm}+2\sum_{i=1}^{3}x_{i}\frac{(x_{i}^{\prime}-x_{i})}{\left|\mathbf{x}-\mathbf{x}^{\prime}\right|^{3}}
  \Biggr\}.\label{eq:mu_0_3}
\end{eqnarray}

To obtain $\mu(\mathbf{a})$ from $\mu(\mathbf{0})$ all that is required
is to make the replacements $\mathbf{x}\rightarrow\mathbf{x}-\mathbf{a}$,
$\mathbf{x}^{\prime}\rightarrow\mathbf{x}^{\prime}-\mathbf{a}$ inside
the braces in Eq.~(\ref{eq:mu_0_3}). The result is
\begin{eqnarray}
  \mu(\mathbf{a}) & = & \frac{G}{2\mathcal{E}}\int d^{3}\mathbf{x}\, d^{3}\mathbf{x}^{\prime}\,\rho(\mathbf{x})
  \rho(\mathbf{x}^{\prime})\times\nonumber \\
  &  & \hspace{0.7cm}\Biggl\{4\pi\delta^{3}\left(\mathbf{x}-\mathbf{x}^{\prime}\right)\left|\mathbf{x}-\mathbf{a}\right|
  ^{2}+\sum_{i,k=1}^{3}(x_{i}-a_{i})(x_{k}-a_{k})\frac{3(x_{i}^{\prime}-x_{i})(x_{k}^{\prime}-x_{k})-\left|\mathbf{x}-
      \mathbf{x}^{\prime}\right|^{2}\delta_{ik}}{\left|\mathbf{x}-\mathbf{x}^{\prime}\right|^{5}}\nonumber \\
  &  & \hspace{0.85cm}+2\sum_{i=1}^{3}(x_{i}-a_{i})\frac{(x_{i}^{\prime}-x_{i})}{\left|\mathbf{x}-\mathbf{x}^{\prime}
    \right|^{3}}\Biggr\}.\label{eq:mu_a}
\end{eqnarray}

The extremization with respect to $\mathbf{a}$ of this quadratic
expression in $\mathbf{a}$ is straightforward, and leads to a $3\times3$
linear problem,
\begin{equation}
\mathbf{Ma}=\mathbf{b},\label{eq:Linear}
\end{equation}
with
\begin{eqnarray}
  \left[\mathbf{M}\right]_{ik} & \equiv & \int d^{3}\mathbf{x}\, d^{3}\mathbf{x}^{\prime}\,\rho(\mathbf{x})\rho(\mathbf{x}
  ^{\prime})\times\left\{ 4\pi\delta^{3}\left(\mathbf{x}-\mathbf{x}^{\prime}\right)\delta_{ik}+\frac{3(x_{i}-x_{i}^{\prime})
      (x_{k}-x_{k}^{\prime})-\left|\mathbf{x}-\mathbf{x}^{\prime}\right|^{2}\delta_{ik}}{\left|\mathbf{x}-\mathbf{x}^{\prime}
      \right|^{5}}\right\} \nonumber \\
  & = & \int d^{3}\mathbf{x}\, d^{3}\mathbf{x}^{\prime}\,\rho(\mathbf{x})\rho(\mathbf{x}^{\prime})\times\left\{ 4\pi
    \delta^{3}\left(\mathbf{x}-\mathbf{x}^{\prime}\right)\delta_{ik}+\frac{\partial^{2}}{\partial x_{i}\partial x_{k}}\frac{1}
    {\left|\mathbf{x}-\mathbf{x}^{\prime}\right|}\right\} \nonumber \\
  & = & \int d^{3}\mathbf{x}\,\left\{ 4\pi\rho(\mathbf{x})^{2}\delta_{ik}-\rho(\mathbf{x})\, 
    G^{-1}\frac{\partial^{2}\Phi(\mathbf{x})}{\partial x_{i}\partial x_{k}}\right\} ,\label{eq:Matrix}
\end{eqnarray}
and
\begin{eqnarray}
  \left[\mathbf{b}\right]_{i} & = & \int d^{3}\mathbf{x}\, d^{3}\mathbf{x}^{\prime}\,\rho(\mathbf{x})\rho(\mathbf{x}
  ^{\prime})\times\left\{ 4\pi\delta^{3}\left(\mathbf{x}-\mathbf{x}^{\prime}\right)x_{i}+\sum_{k=1}^{3}x_{k}\frac{3(x_{i}-
      x_{i}^{\prime})(x_{k}-x_{k}^{\prime})-\left|\mathbf{x}-\mathbf{x}^{\prime}\right|^{2}\delta_{ik}}{\left|\mathbf{x}-
        \mathbf{x}^{\prime}\right|^{5}}-\frac{x_{i}-x_{i}^{\prime}}{\left|\mathbf{x}-\mathbf{x}^{\prime}\right|^{3}}\right\} 
  \nonumber \\
  & = & \int d^{3}\mathbf{x}\, d^{3}\mathbf{x}^{\prime}\,\rho(\mathbf{x})\rho(\mathbf{x}^{\prime})\times\left\{ 4\pi
    \delta^{3}\left(\mathbf{x}-\mathbf{x}^{\prime}\right)x_{i}+\sum_{k=1}^{3}x_{k}\frac{\partial^{2}}{\partial x_{i}\partial 
      x_{k}}\frac{1}{\left|\mathbf{x}-\mathbf{x}^{\prime}\right|}+\frac{\partial}{\partial x_{i}}\frac{1}{\left|\mathbf{x}-
        \mathbf{x}^{\prime}\right|}\right\} \nonumber \\
  & = & \int d^{3}\mathbf{x}\,\left\{ 4\pi\rho(\mathbf{x})^{2}x_{i}-\rho(\mathbf{x})\sum_{k=1}^{3}x_{k}
    G^{-1}\frac{\partial^{2}\Phi(\mathbf{x})}{\partial x_{i}\partial x_{k}}-\rho(\mathbf{x})G^{-1}\frac{\partial
      \Phi(\mathbf{x})}{\partial x_{i}}\right\} .\label{eq:RHS_1}
\end{eqnarray}
The last term in Eq.~(\ref{eq:RHS_1}) yields, upon integration,
$G^{-1}$ times the net self-force of the gravitating mass configuration,
which is necessarily zero. We therefore have for $\mathbf{b}$
\begin{equation}
  \left[\mathbf{b}\right]_{i}=\int d^{3}\mathbf{x}\,\left\{ 4\pi\rho(\mathbf{x})^{2}x_{i}-\rho(\mathbf{x})\sum_{k=1}^{3}
    x_{k}G^{-1}\frac{\partial^{2}\Phi(\mathbf{x})}{\partial x_{i}\partial x_{k}}\right\} .\label{eq:RHS}
\end{equation}

The integrands in Eqs. (\ref{eq:Matrix}) and (\ref{eq:RHS}) feature
the sum of a term proportional to the square of the density, and a
term proportional to the tidal tensor $\partial^{2}\Phi/\partial x_{i}\partial x_{k}$.
In the next section, we estimate the relative sizes of the two terms in
each of the two integrals, and find that it is an acceptable approximation
to drop the tidal terms in comparison with the square-density terms.
Making this approximation, we obtain
\begin{eqnarray}
  \mathbf{M} & \approx & \mathbf{1}\times\int d^{3}\mathbf{x}\,4\pi\rho(\mathbf{x})^{2}\label{eq:Matrix_Approx}\\
  \mathbf{b} & \approx & \int d^{3}\mathbf{x}\,4\pi\rho(\mathbf{x})^{2}\,\mathbf{x},\label{eq:RHS_Approx}
\end{eqnarray}
so that
\begin{eqnarray}
  \mathbf{a} & \approx & \frac{\int d^{3}\mathbf{x}\,\mathbf{x}\rho(\mathbf{x})^{2}}{\int d^{3}\mathbf{x}\,
    \rho(\mathbf{x})^{2}}\nonumber \\
  & \equiv & \left\langle \mathbf{x}\right\rangle _{\rho^{2}}.\label{eq:a_approx_again}
\end{eqnarray}
That is to say, the optimal expansion center location is the average
location weighted by the square of the density.

A useful result worth setting down is a formula for the spectral compactness,
$\mu(\mathbf{0})$ that is convenient for numerical computation. Starting
from Eq.~(\ref{eq:mu_0_3}), we may replace the dipole and quadrupole
tensors with suitable derivatives of the Green's function, as we did
in Eqs.~(\ref{eq:Matrix}) and (\ref{eq:RHS_1}). We find that 
\begin{equation}
  \mu(\mathbf{0})=\frac{G}{2\mathcal{E}}\int d^{3}\mathbf{x}\,\left\{ 4\pi\left|\mathbf{x}\right|^{2}\rho(\mathbf{x})^{2}+
    \rho(\mathbf{x})G^{-1}\sum_{i,k=1}^{3}x_{i}x_{k}\frac{\partial^{2}\Phi(\mathbf{x})}{\partial x_{i}\partial 
      x_{k}}-2\rho(\mathbf{x})G^{-1}\sum_{i=1}^{3}x_{i}\frac{\partial\Phi(\mathbf{x})}{\partial x_{i}}\right\} .
\label{eq:Compactness_Formula}
\end{equation}
All the data required to compute this integral over the domain is
available after the potential has been computed.

\section{Estimating Spectral Compactness Integrals}

In this appendix we estimate the relative sizes of the two terms in
Eqs.~(\ref{eq:Matrix}) and (\ref{eq:RHS}).

By re-expressing the potential $\Phi(\mathbf{x})$ in terms of the
density $\rho(\mathbf{x}),$ and making the change of variables
$\mathbf{x}^{\prime}\rightarrow\mathbf{y}=\mathbf{x}
^{\prime}-\mathbf{x}$, Eqs.~(\ref{eq:Matrix}) and (\ref{eq:RHS}) may
be written as
\begin{eqnarray}
  \left[\mathbf{M}\right]_{ik} & = & \int d^{3}\mathbf{x}\,\rho(\mathbf{x})\,\int d^{3}\mathbf{y}\rho(\mathbf{x}+\mathbf{y})
  \left\{ 4\pi\delta^{3}\left(\mathbf{y}\right)\delta_{ik}-\frac{y^{2}\delta_{ik}-3y_{i}y_{k}}{y^{5}}\right\} 
  \label{eq:Matrix_Rel}\\
  \left[\mathbf{b}\right]_{i} & = & \int d^{3}\mathbf{x}\,\rho(\mathbf{x})\,\int d^{3}\mathbf{y}\rho(\mathbf{x}+\mathbf{y})
  \left\{ 4\pi\delta^{3}\left(\mathbf{y}\right)x_{i}-\frac{\sum_{k=1}^{3}x_{k}\left[y^{2}\delta_{ik}-3y_{i}y_{k}\right]}
    {y^{5}}\right\} .\label{eq:RHS_rel}
\end{eqnarray}
We single out the rational terms in these integrals:
\begin{eqnarray}
  \left[\mathbf{M}^{R}\right]_{ik} & \equiv & -\int d^{3}\mathbf{x}\,\rho(\mathbf{x})\,\int d^{3}\mathbf{y}\rho(\mathbf{x}+
  \mathbf{y})\frac{y^{2}\delta_{ik}-3y_{i}y_{k}}{y^{5}}\label{eq:Matrix_rat}\\
  \left[\mathbf{b}^{R}\right]_{i} & \equiv & -\int d^{3}\mathbf{x}\,\rho(\mathbf{x})\,\int d^{3}\mathbf{y}\rho(\mathbf{x}+
  \mathbf{y})\frac{\sum_{k=1}^{3}x_{k}\left[y^{2}\delta_{ik}-3y_{i}y_{k}\right]}{y^{5}}.\label{eq:RHS_rat}
\end{eqnarray}

The numerators of the integrands contain the trace-free symmetric
tensor $y^{2}\delta_{ik}-3y_{i}y_{k}$. We recognize this as quadrupole
tensor, and exploit its nature as a spherical tensor --- a spherical
harmonic in tensor guise --- to reduce the order of the singularity
in $y$.

We will require the following spherical integrals:
\begin{eqnarray}
\oint d^{2}\mathbf{n}\, n_{i}n_{k} & = & \frac{4\pi}{3}\delta_{ik}\label{eq:oint_2}\\
\oint d^{2}\mathbf{n}\, n_{i}n_{k}n_{l}n_{m} & = & \frac{4\pi}{15}\left(\delta_{ik}\delta_{ml}+\delta_{im}\delta_{kl}+
\delta_{il}\delta_{mk}\right).\label{eq:oint_4}
\end{eqnarray}
These may be obtained by observing that the resulting tensors must
be rotationally-invariant and totally symmetric under index interchange.
Such tensors can only be constructed from the only tensor at hand
--- the identity tensor $\delta_{ik}$ --- by the combinations indicated.
The coefficients may then be calculated by setting $i=k=l=m=3$ in
the resulting expressions and performing the integrals in spherical
coordinates. In addition, we observe that any similar integral featuring
an odd number of components of $\mathbf{n}$ as factors in the integrand
is necessarily zero, since it changes sign under the variable change
$\mathbf{n}\rightarrow-\mathbf{n}$.$ $ Note also that Eq.~(\ref{eq:oint_2})
implies that the spherical integral of the quadrupole tensor $\delta_{ik}-3n_{i}n_{k}$
is zero.

Since we are interested in the $y\rightarrow0$ behavior, we expand
$\rho(\mathbf{x}+\mathbf{y})$ around $\mathbf{x}$:
\begin{equation}
  \rho(\mathbf{x}+\mathbf{y})=\rho(\mathbf{x})+\sum_{l=1}^{3}y_{l}\frac{\partial\rho(\mathbf{x})}{\partial x_{l}}+\frac{1}
  {2}\sum_{l=1}^{3}\sum_{m=1}^{3}\left[y_{l}y_{m}+\mathcal{O}(y^{3})\right]\frac{\partial^{2}\rho(\mathbf{x})}{\partial 
    x_{l}\partial x_{m}}.\label{eq:Expand_rho}
\end{equation}

Inserting this expansion in Eq.~(\ref{eq:Matrix_rat}), we obtain
\begin{eqnarray}
  \left[\mathbf{M}^{R}\right]_{ik} & \approx & -\int d^{3}\mathbf{x}\,\rho(\mathbf{x})\,\frac{1}{2}\sum_{l=1}^{3}\sum_{m=1}
  ^{3}\frac{\partial^{2}\rho(\mathbf{x})}{\partial x_{l}\partial x_{m}}\int_{0}^{\infty}y^{2}dy\,\oint d^{2}\mathbf{n}\,
  \left[y_{l}y_{m}+\mathcal{O}(y^{3})\right]\frac{y^{2}\delta_{ik}-3y_{i}y_{k}}{y^{5}}\nonumber \\
  & = & -\int d^{3}\mathbf{x}\,\rho(\mathbf{x})\,\frac{1}{2}\sum_{l=1}^{3}\sum_{m=1}^{3}\frac{\partial^{2}\rho(\mathbf{x})}
  {\partial x_{l}\partial x_{m}}\int_{0}^{\infty}\left[y+\mathcal{O}(y^{2})\right]\, dy\oint d^{2}\mathbf{n}
  \left(\delta_{ik}n_{l}n_{m}-3n_{i}n_{k}n_{l}n_{m}\right)\nonumber \\
  & = & -\int d^{3}\mathbf{x}\,\rho(\mathbf{x})\,\frac{1}{2}\sum_{l=1}^{3}\sum_{m=1}^{3}\frac{\partial^{2}\rho(\mathbf{x})}
  {\partial x_{l}\partial x_{m}}\int_{0}^{\infty}\left[y+\mathcal{O}(y^{2})\right]\, dy\,\frac{4\pi}{15}\left(2\delta_{ik}
    \delta_{lm}-3\delta_{il}\delta_{km}-3\delta_{im}\delta_{kl}\right)\nonumber \\
  & = & -\frac{4\pi}{15}\int d^{3}\mathbf{x}\,\rho(\mathbf{x})\,\left(\delta_{ik}
    \nabla^{2}\rho(\mathbf{x})-3\frac{\partial^{2}\rho(\mathbf{x})}{\partial x_{i}\partial x_{k}}\right)\int_{0}^{\infty}
  \left[y+\mathcal{O}(y^{2})\right]\, dy,\label{eq:Matrix_limit}
\end{eqnarray}
where in the first line we summarily dropped from the density expansion
both the $\mathcal{O}(y^{0})$ term --- because it results in a spherical
integral of the quadrupole tensor, which is zero --- and the $\mathcal{O}(y^{1})$
term --- because it results in a spherical integral with an odd number
of vector factors, which is also zero.
In the inner integrand, we see that the dependence on $y$ as $y\rightarrow0$
is a very benign $\mathcal{O}(y^{1})$.

We proceed similarly, inserting the expansion of Eq.~(\ref{eq:Expand_rho})
into Eq.~(\ref{eq:RHS_rat}). Again, only one term from the expansion
survives, with the $\mathcal{O}(y^{2})$ term latching on to the quadrupole,
as before. We obtain
\begin{equation}
  \left[\mathbf{b}^{R}\right]_{i}\approx-\frac{4\pi}{15}\int d^{3}\mathbf{x}\,\rho(\mathbf{x})\,\left\{ \left(x_{i}
      \nabla^{2}\rho(\mathbf{x})-3\sum_{k=1}^{3}x_{k}\frac{\partial^{2}\rho(\mathbf{x})}{\partial x_{l}\partial x_{k}}\right)
  \right\} \int_{0}^{\infty}\left[y+\mathcal{O}(y^{2})\right]\, dy.\label{eq:RHS_limit}
\end{equation}

We again find that the ``singular'' behavior of the integrand is
in fact $\mathcal{O}(y^{1})$ as $y\rightarrow0$. This $\mathcal{O}(y^{1})$
behavior is no different from the short-distance behavior of the Poisson
Green's function, which combines a $y^{-1}$ singularity with the $d^{3}\mathbf{y}$
measure to produce an $\mathcal{O}(y^{1})$ dependence in the integrand.

We now use the expressions just derived to estimate the relative size
of the rational terms and the $\delta$-function terms in
Eqs.~(\ref{eq:Matrix_Rel}) and (\ref{eq:RHS_rel}). To do this, we
assume a distribution of matter bounded to some region of size $R$. We
estimate the term $\int dy[y+\mathcal{O}(y^{2})]\sim R^{2}/2$.  We
also assume the presence of a sharp peak in the density, so that the
integral measure $d^{3}\mathbf{x}\,\rho(\mathbf{x})$ places most of
the action near the density peak. In this region, the linear term in
the expansion in Eq.~(\ref{eq:Expand_rho}) is small compared to the
$\mathcal{O}(y^{2})$ term, and may be neglected, and we may estimate
$\frac{1}{2}R^{2}|\partial^{2}\rho/\partial x^{2}|\sim\rho$.  We use
this estimate for the aggregate second-derivative terms in brackets in
Eqs.~(\ref{eq:Matrix_limit}) and (\ref{eq:RHS_limit}).  We also
replace the factors $x$ in Eq.~(\ref{eq:RHS_limit}) by a typical value
$R$. By these means, we obtain for the size of the matrix elements of
$\mathbf{M}^{R}$
\begin{eqnarray}
  \left|\left[\mathbf{M}^{R}\right]_{ik}\right| & \sim & \frac{4\pi}{15}\int d^{3}\mathbf{x}\,\rho(\mathbf{x})^{2}.
\label{eq:Estimate_1}
\end{eqnarray}

Since the $\delta$-function term in Eq.~(\ref{eq:Matrix_Rel}) is
\begin{equation}
  \left[\mathbf{M}^{\delta}\right]_{ik}=4\pi\delta_{ik}\int d^{3}\mathbf{x}\,\rho(\mathbf{x})^{2},\label{eq:M_delta}
\end{equation}
we obtain the ratio
\begin{equation}
  \left|\frac{\left[\mathbf{M}^{R}\right]_{ik}}{\left[\mathbf{M}^{\delta}\right]_{ik}}\right|\sim\frac{1}{15},
\label{eq:Estimate_1_ratio}
\end{equation}
give or take a little slop. By the same means, we obtain
\begin{equation}
  \left|\left[\mathbf{b}^{R}\right]_{k}\right|\sim\frac{4\pi R}{15}\int d^{3}\mathbf{x}\,\rho(\mathbf{x})^{2}.
\label{eq:Estimate_2}
\end{equation}

The $\delta$-function term in Eq.~(\ref{eq:RHS_rel}) is
\begin{eqnarray}
  \left[\mathbf{b}^{\delta}\right]_{k} & = & 4\pi\int d^{3}\mathbf{x}\,\rho(\mathbf{x})^{2}\, x_{k}\nonumber \\
  & \sim & 4\pi R\int d^{3}\mathbf{x}\,\rho(\mathbf{x})^{2},\label{eq:b_delta}
\end{eqnarray}
so that the ratio of terms is again
\begin{equation}
  \left|\frac{\left[\mathbf{b}^{R}\right]_{k}}{\left[\mathbf{b}^{\delta}\right]_{k}}\right|\sim\frac{1}{15}.
\label{eq:Estimate_2_ratio}
\end{equation}

On the basis of these estimates it appears that the neglect of the
rational terms in Eqs.~(\ref{eq:Matrix}) and (\ref{eq:RHS}) is
a justifiable approximation.

\bibliography{papersDB}

\end{document}